\documentclass[prl, aps, twocolumn, superscriptaddress,noshowpacs]{revtex4}
\usepackage{eqnarray}
\usepackage[utf8]{inputenc}
\usepackage[english]{babel} 
\usepackage[english]{layout}
\usepackage{amsmath,amsfonts,amssymb}
\usepackage{graphicx}
\usepackage{float}
\usepackage{color}
\usepackage{braket}
\usepackage{version}
\usepackage{array,multirow,makecell}

\begin{document}

\title{Analog Kerr Black hole and Penrose effect in a Bose-Einstein Condensate}

\author{D. D. Solnyshkov}
\affiliation{Institut Pascal, University Clermont Auvergne, CNRS, SIGMA Clermont, F-63000 Clermont-Ferrand, France} 
\author{C. Leblanc}
\affiliation{Institut Pascal, University Clermont Auvergne, CNRS, SIGMA Clermont, F-63000 Clermont-Ferrand, France} 
\author{S. V. Koniakhin}
\affiliation{Institut Pascal, University Clermont Auvergne, CNRS, SIGMA Clermont, F-63000 Clermont-Ferrand, France} 
\affiliation{St. Petersburg Academic University - Nanotechnology Research and Education Centre of the Russian Academy of Sciences, 194021 St. Petersburg, Russia}
\author{O. Bleu}
\affiliation{Institut Pascal, University Clermont Auvergne, CNRS, SIGMA Clermont, F-63000 Clermont-Ferrand, France} 
\author{G. Malpuech}
\affiliation{Institut Pascal, University Clermont Auvergne, CNRS, SIGMA Clermont, F-63000 Clermont-Ferrand, France} 

\begin{abstract}
Analog physics allows simulating inaccessible objects, such as black holes, in the lab. We propose to implement an acoustic Kerr black hole with quantized angular momentum in a polariton Bose-Einstein condensate. We show that the metric of the condensate is equivalent to the Kerr's one, exhibiting a horizon and an ergosphere. Using topological defects as test particles, we demonstrate an analog Penrose effect, extracting the rotation energy of the black hole. The particle trajectories are well described by the time-like geodesics of the Kerr metric, confirming the potential of analog gravity.
\end{abstract}
\maketitle

The field of analog physics has been growing since the 80s, when the first proposals were published, linking astrophysical phenomena, such as the Hawking emission of the black holes \cite{Hawking1974} or the Kibble mechanism of the topological defect formation in the early Universe \cite{Kibble1976} with desktop systems \cite{Unruh1981,Zurek1985,VolovikReview}. The long efforts \cite{Barcelo2005} were crowned with a great success, allowing the recent observation of Hawking emission \cite{Weinfurtner2011,Steinhauer2014} and the observation of  the Kibble-Zurek mechanism in various systems \cite{Bowick1994,Feng2018}. Of course, the full spectrum of analog physics phenomena extends far beyond these two most striking examples \cite{Bloch2012,Faccio2013,Armitage2018,Eckel2018}.

Exciton-polariton (polariton) condensates are particularly well suited for analog physics because this quantum fluid allows a versatile all-optical control of the wavefunction via external potentials (both real and imaginary), together with the optical means for the measurement of all wavefunction components both in real and in reciprocal space. Polaritons are light-matter quasiparticles existing in microcavities\cite{Microcavities} in the strong coupling regime. They inherit a small effective mass of photons and strong interparticle interactions from excitons, the former ensuring a large coherence length even in reduced dimensions \cite{Wertz2010}, and the latter providing the means for optical potential engineering \cite{Sanvitto2011}. Many proposals for analog physics based on polaritons have appeared in the recent years \cite{Solnyshkov2011,Gerace2012,Solnyshkov2012,Solnyshkov2016}, and some of them have already been implemented experimentally \cite{Hivet,Nguyen2015}. 

A new frontier in the analog physics is the Penrose effect \cite{Penrose1971}: the extraction of kinetic energy stored in rotating black holes described by the Kerr metric \cite{Kerr1963}. It involves the creation of a pair of positive and negative energy particles, when the particle with negative energy $E<0$ falls into the black hole, reducing its angular momentum, while the particle with positive energy $E>0$ escapes to infinity, with the overall process resulting in the energy extraction. The region where such process is possible is called an ergosphere. As a first step, it requires the creation of a Kerr black hole analog in a fluid \cite{Visser1998,Vocke2018}, either classical or quantum.
So far, the acoustic black holes have been mostly 1D \cite{Lahav2010,Steinhauer2014,Nguyen2015,Euve2016}, and only recently 2D black holes with closed horizons have been implemented, which has already allowed to observe the superradiance effect \cite{Torres2017}. However, only the propagation of sound waves corresponding to the null geodesics has been studied so far. Little attention has been paid in this respect to quantum vortices, which represent an important direction of analog physics on their own (emergent electrodynamics) since 1970s\cite{Popov1972,Ambegaokar1980,Lundh2000,Zhang2004}. They behave as relativistic particles with their dynamics governed by the fluid metric \cite{Zhang2004,VolovikReview}. Contrary to high-wavevector bogolons,  which propagate at velocities larger than the speed of sound in the condensate $c$ and thus violate the horizons of the low-wavevector metric, vortices cannot exceed $c$ \cite{Pitaevskii}. With their mass given by the Einstein's relation $E=mc^2$ \cite{Popov1972,Duan1994}, and their stability ensured by a topological quantum number (winding) \cite{Thouless1998}, 
vortices therefore appear as excellent candidates to study the time-like geodesics of massive particles in vicinity of black holes. The dynamics of vortices has been extensively studied experimentally in atomic \cite{Ketterle2001,Freilich2010} and polariton condensates \cite{Nardin2011,Dominici2018}, but not in the framework of analog physics.

In this work, we propose to implement a Kerr black hole in a polariton condensate by combining optical excitation with a Gauss-Laguerre beam, providing the angular momentum, and a region of reduced lifetime, creating an inward flow. Similar configurations have already been implemented with polaritons \cite{Boulier2015,Ohadi2016}. We show that the metric of the condensate in this configuration is equivalent to the Kerr metric of a rotating black hole. We demonstrate that the topological defects of the condensate (quantum vortices) can be used as test particles whose propagation follows the time-like geodesics of the Kerr metric, opening the domain of analog gravity to the studies of massive particles. We simulate the Penrose effect using a vortex-antivortex pair, with an antivortex falling into the black hole and reducing its angular momentum, and a vortex escaping from the black hole to the infinity. This represents a first step towards self-consistent analog gravity systems with the metric naturally subject to quantum fluctuations. While we have optimized our proposal for the cavity polariton system, there are no fundamental obstacles for its implementation in other other types of quantum fluids, such as atomic condensates \cite{Takeuchi2008} or superfluid light \cite{Michel2018}.

A Bose-Einstein condensate is a quantum fluid which can be described in the mean-field approximation by the Gross-Pitaevskii equation \cite{Pitaevskii}:
\begin{equation}
i\hbar \frac{{\partial \psi }}{{\partial t}} =  - \frac{{{\hbar ^2}}}{{2m}}\Delta \psi  + \alpha {\left| \psi  \right|^2}\psi +U\psi - \mu \psi 
\label{GPE}
\end{equation}
where $\psi(r,t)$ is the condensate wavefunction, $m$ is the particle mass, $\alpha$ is the interaction constant, $U$ is the external potential (with a possible imaginary part describing particle decay), and $\mu$ is the chemical potential. The analogy between the condensate and the general relativity spacetime is based on the fact that the propagation of the weak excitations of a homogeneous stationary condensate can be described by a relativistic wave equation for their phase \cite{Garay2000}:
\begin{equation}
\partial_{\nu}(\sqrt{-g}g^{\mu\nu}\partial_{\nu}\varphi)=0
\end{equation}
with $g=det(g_{\mu\nu})$, and the metric $g_{\mu\nu}$ totally determined by the background stationary velocity $\textbf{v}=(\hbar/m)\nabla\arg\psi$ and the local speed of sound $c=\sqrt{\alpha |\psi|^2/m}$, being in the most general case given by \cite{Visser1998}: 
\begin{equation}
g_{\mu\nu}=\frac{mn}{c}\begin{pmatrix} 
-(c^{2}-\textbf{v}^{2}) & \vdots & -\textbf{v}\\
 \ldots \ldots \ldots & . & \ldots \ldots \\
-\textbf{v} & \vdots & \delta_{ij}\\
 \end{pmatrix} \label{gmetric}
\end{equation}
Derived for weak density waves, this metric is also obeyed by vortices \cite{Zhang2004}.

The Kerr metric of a rotating black hole (in the equatorial plane only, since a 2D condensate can reproduce only a single plane) can be written in Boyer–Lindquist coordinates as
\begin{equation}
\label{Kerr}
g_{\mu\nu}^{Kerr} =\begin{pmatrix} 
-\bigg(1-\frac{2M}{r}\bigg) & 0 & -\frac{4aM}{r}\\
0 & \frac{r^2}{r^2-2Mr+a^2} & 0\\
-\frac{4aM}{r} & 0 & \bigg(r^2+a^2+\frac{2a^2M}{r}\bigg)
\end{pmatrix}
\end{equation}
where $a$ is the black hole angular momentum.

To reproduce such a metric, we consider a cylindrically-symmetric configuration with radial and azimuthal flows ($v_r$, $v_\phi$), with an appropriate change of coordinates \cite{Berti2004}, which allows to write the metric of the condensate as:
\begin{equation}
\label{Kcm}
g_{\mu\nu}=\frac{mn}{c}
\begin{pmatrix} 
-(c^{2}-{v}^{2}) & 0 & -2rv_\phi\\
0 & \bigg(1-\frac{v_r^2}{c^2}\bigg)^{-1} & 0 \\
-2rv_\phi & 0 & r^2\\
 \end{pmatrix}
 \end{equation}
Here, the flow velocity $\mathbf{v}(\mathbf{r})$ and the speed of sound $c(\mathbf{r})$ are both functions of coordinates, determined by the unperturbed condensate wavefunction $\psi$. In order to obtain an analog of a Kerr black hole, we need to generate a configuration where these functions would have a proper behavior. Quantum fluids are irrotational, and their azimuthal flow is controlled by the number $\nu$ of quantum vortices which are topological defects, characterized by the quantized circulation of angular velocity. On the other hand, a radial flow requires a sink (drain, particle decay) in the central region. Both can be combined using existing techniques \cite{Boulier2015} in polariton condensates: a macroscopically occupied state can be created or seeded by a Gauss-Laguerre beam carrying required angular momentum \cite{Marchetti2010} (in presence of a $cw$ non-resonant or quasi-resonant pumping), and a shorter lifetime can be provided by a localized defect in the cavity mirrors (or a $\mu$m size metal deposit) on which the beam should be centered. 

Since even for a single vortex a complete analytical solution of the Gross-Pitaevskii equation has not been found yet, we use an asymptotic series expansion at $r\gg\xi$ in order to find $\psi$ , where $\xi=\hbar/\sqrt{2\alpha n m}$ is the size of a vortex core (the healing length). In this approximation, both the sink and the vorticity concentrated in the central region ($r\ll\xi$) can be approximated as delta functions:
\begin{eqnarray}
\nabla\times\mathbf{v}&=&2\pi\nu\frac{\hbar}{m}\delta_{2D}(\mathbf{r})\\
\nabla\cdot\mathbf{v}&=&-2\pi\zeta\frac{\hbar}{m}\delta_{2D}(\mathbf{r})\nonumber
\end{eqnarray}
where $\nu$ is an integer determining vorticity and $\zeta>0$ is obtained \emph{a posteriori} from particle decay in the center. This allows to find the components of the velocity: $v_\phi=\hbar\nu/mr$, $v_r=\hbar\zeta/mr$
and the approximate solution for the wavefunction of the condensate:
\begin{equation}
\label{wavef}
\psi(r,\phi)=\sqrt{n_\infty}\left(1-\xi^2\frac{\nu^2+\zeta^2}{r^2}\right)\exp\left(i\left(\zeta\ln\frac{r}{\xi}+\nu\phi\right)\right)
\end{equation}
The scale of density variation is increased by $\sqrt{\nu^2+\xi^2}$, ensuring a relatively slow metric variation for test waves and particles.
Then, we find the analytical expressions for the radius of the event horizon $r_h$ and the radius of the ergosphere (the static limit) $r_s$. Indeed, the event horizon is determined \cite{Visser2005} by the change of sign of the metric component $g_{rr}$ ($v_r=c$), which gives:
\begin{equation} 
r_{H}=\frac{\xi}{\sqrt{2}}\left(\zeta+\sqrt{3\zeta^2+2\nu^2}\right)
\label{rhori}
\end{equation}
while the static limit is determined by the change of sign of $g_{tt}$ ($v=c$) which gives
\begin{equation}
r_{E}=\frac{1+\sqrt{3}}{\sqrt{2}}\xi\sqrt{\zeta^2+\nu^2}
\end{equation}
Same as the condensate wavefunction \eqref{wavef}, both expressions are only valid if $r_{H,E}\gg\xi$.

First, we check that the proposed configuration is realistic and that the analytical solution is correct. In order to find the stationary solution for the condensate wavefunction containing $\nu=16$ vortices in the short-lifetime region, 
we solve the Gross-Pitaevskii equation numerically with a relaxation term \cite{Pitaevskii58}, using the typical parameters for GaAs cavities ($\alpha=5~\mu$eV$\mu$m$^2$, $m=5\times10^{-5}m_0$ \cite{Ferrier2011}, see \cite{suppl} for more details). The results are shown in Fig. 1. As expected, the initial single-vortex state with high angular momentum splits \cite{Castin1999,Kawaguchi2004} into $\nu$ single-charged vortices kept inside the horizon by the convergent flow \cite{suppl}. 
 Panel (a) shows the density profiles: the analytical approximation with $\zeta\approx 7$ (red dash-dotted) gives a good fit to numerics (black solid) for $r>16~\mu$m  and the calculated positions of the event horizon (black dashed) and the static limit (blue dashed) are within the region of validity of the approximation. Their correctness is confirmed by the numerical velocities (panel b), their position given by the crossing of the speed of sound $c$ (red) with the radial velocity $v_r$ for the horizon (black) and total velocity $v$ for the static limit (blue), with the ergoregion contained between the two. It is interesting to note that the maximal angular momentum of such analog black hole is limited, as for a real black hole ($a/M<1$). For given $\zeta$, the analog black hole cannot contain more than $\nu_{max}=r_H/\xi$ vortices \cite{suppl}.

\begin{figure}[tbp]
  \includegraphics[width=0.49\textwidth]{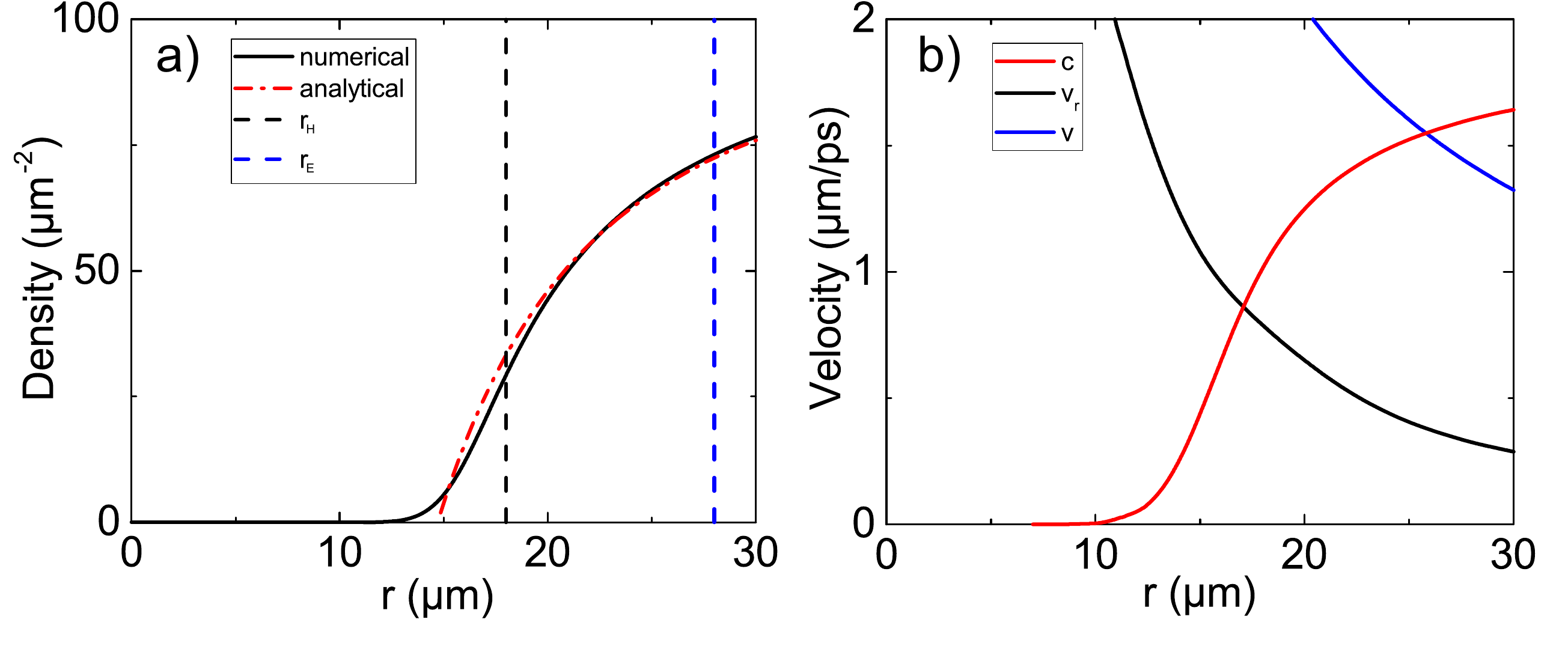}\\
  \caption{(Color online) a) Numerical solution of the Gross Pitaevskii equation (black solid line) and the analytical density profile (red dash-dotted line). Black dashed line -- horizon, blue dashed line -- static limit. b) Velocities determining the metric (from the numerical solution): $c$ (black solid), $v_r$ (blue dashed), $v$ (red dash-dotted).}
  \label{fit}
\end{figure}

\begin{figure}[tbp]
    \centering
    \includegraphics[width=0.49\textwidth]{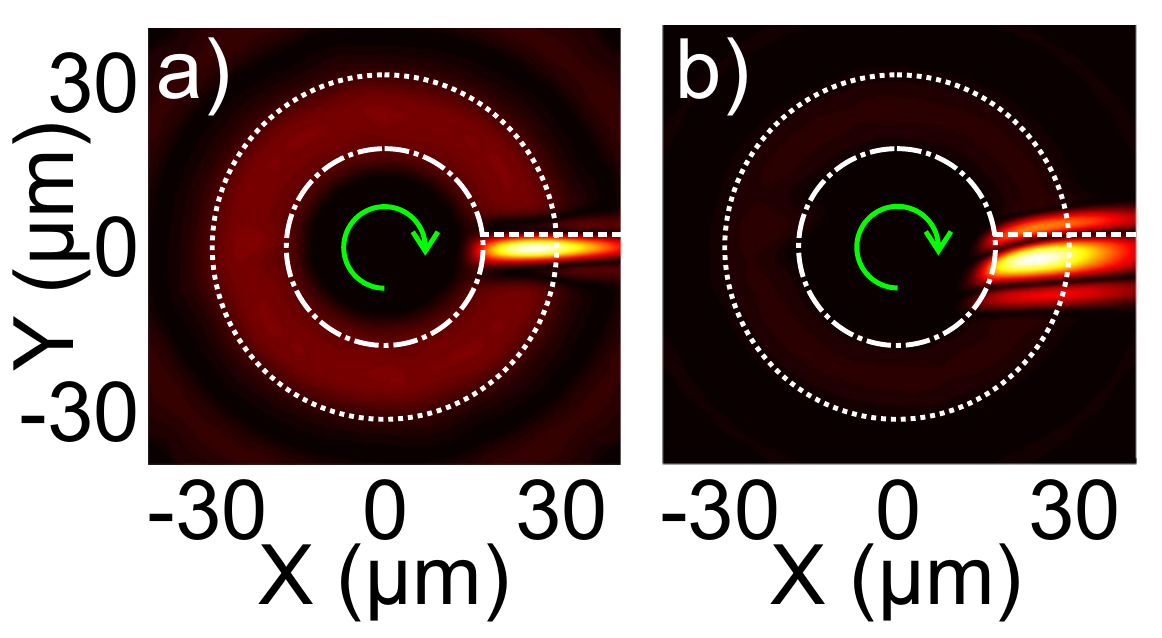}
    \caption{Simulation of an acoustic Kerr black hole with a density wave generated along $X$. Images show the difference between the stationary solution $\psi_0$ and the perturbed solution $\psi(\mathbf{r},t)$: (a) $t=0$ (white dashed line marks the reference node of the density wave), (b) $t=2$ ps (the node of the wave moves down inside the ergosphere and up outside). Dashed circle shows the static limit and dash-dotted -- the horizon. Green arrows mark the rotation direction.}
    \label{fig2}
\end{figure}

Next, we show that the behavior of the weak excitations of the condensate indeed corresponds to the metric \eqref{Kcm}. We solve the Gross-Pitaevskii equation \eqref{GPE}  numerically over time, taking the condensate wavefunction $\psi_0$ found previously as an initial condition, maintaining constant particle density at large distance. Weak density waves are created by a shallow, localized, and short potential pulse with a Gaussian shape. In practice, such potential can be created by a laser pulse. Figure~\ref{fig2} shows two snapshots of the spatial images of the absolute value of the deviation from the stationary configuration $||\psi|^2-|\psi_0|^2|$. Panel (a) shows a perturbation stretched along the X axis, giving rise to two waves with radial velocities $\pm c$ with respect to the fluid. A node in this density wave used as a reference is marked by a dashed white line. Panel (b) shows the evolution of this image after 2 ps. As expected, inside the ergosphere (marked by a white dashed circle) both density waves propagate downwards (in the direction of rotation of the black hole) due to frame dragging, whereas outside both propagation directions are possible.

\begin{figure}[tbp]
  \includegraphics[width=0.49\textwidth]{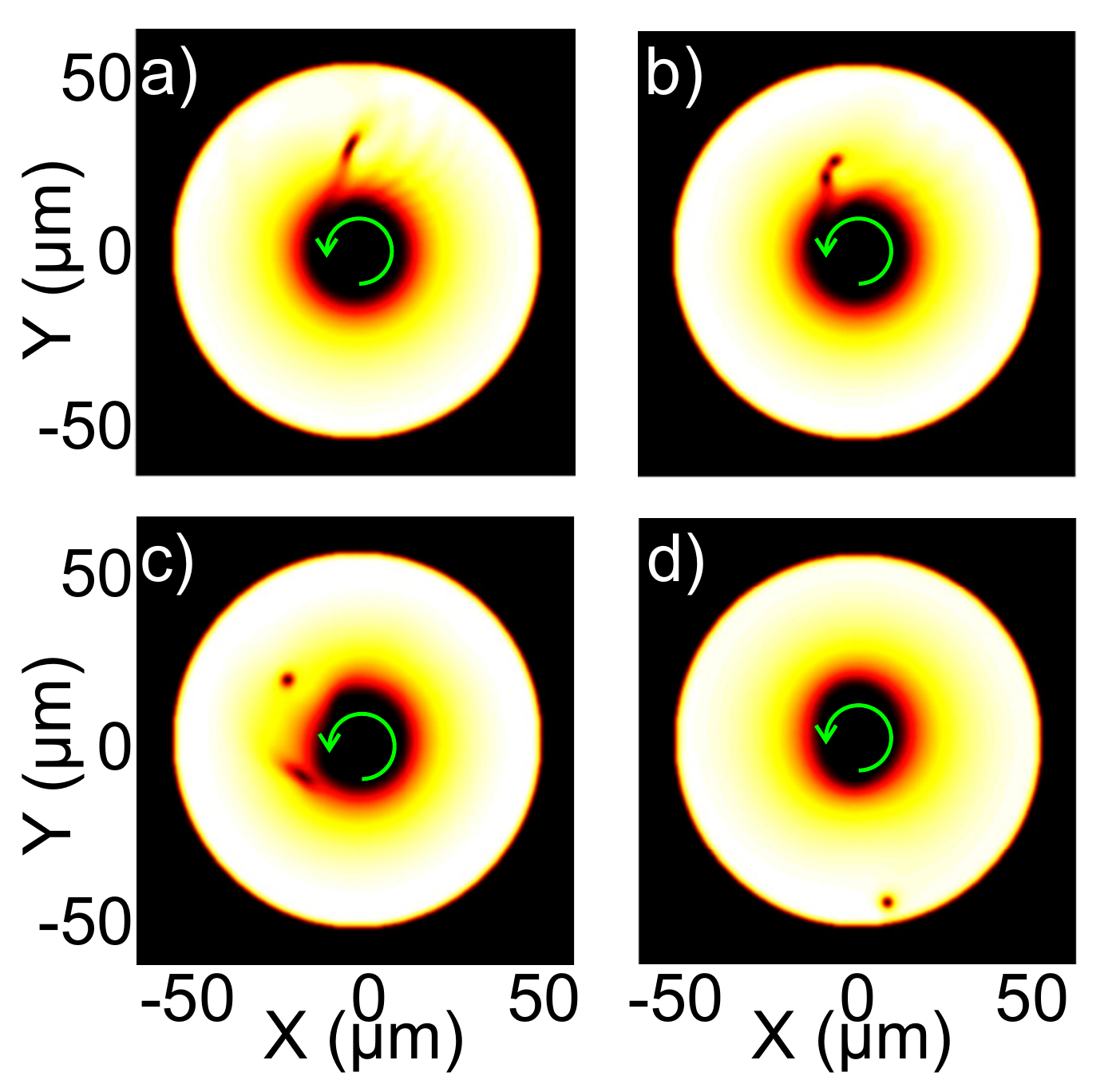}\\
  \caption{(Color online)  Snapshots of the condensate density during the Penrose process: a) creation of a density dip; b) formation of a vortex-antivortex pair; c) annihilation of one of the vortices of the black hole; d) escape of the vortex. Green arrows mark the rotation direction.}
  \label{snapshots}
\end{figure}

Now that we have confirmed that our configuration provides a good analogy with a Kerr black hole for density waves (corresponding to null geodesics), we make the ultimate step, simulating the Penrose process with particles (vortices). The snapshots of the process are shown in Fig.~\ref{snapshots}. We now use a strong localized potential pulse to create a strong density perturbation (Fig.~\ref{snapshots}(a)). In a homogeneous condensate, this localized density dip corresponds to the final stage of a vortex-antivortex pair annihilation \cite{Pitaevskii}, propagating with a speed of sound $c$ and disappearing with time. However, if such density dip is created inside the ergosphere, the opposite effect is observed: the density dip becomes elongated because of the spaghettification \cite{Hawking1988}, and a vortex-antivortex pair appears (Fig.~\ref{snapshots}(b), microscopic mechanism discussed in \cite{suppl}). The antivortex, being closer to the horizon and corresponding to a negative-energy state $E<0$ is rapidly absorbed by the black hole and annihilates a single vortex inside it (Fig.~\ref{snapshots}(c)), reducing its angular momentum. The remaining vortex, depending on the position of its creation, can either escape from the black hole to infinity (the edge of the cavity, Fig.~\ref{snapshots}(d)) or fall into the black hole. The negative energy of the antivortex is confirmed by its velocity \cite{Taylor,suppl} and by the  energy conservation (it allows the vortex to gain enough energy to escape). Indeed, a single vortex created at the same distance at $t=0$ always falls into the black hole \cite{suppl}. We note that the ringdown effect of the perturbed black hole is also visible \cite{Patrick2018}.

\begin{figure}[tbp]
  \includegraphics[width=0.9\linewidth]{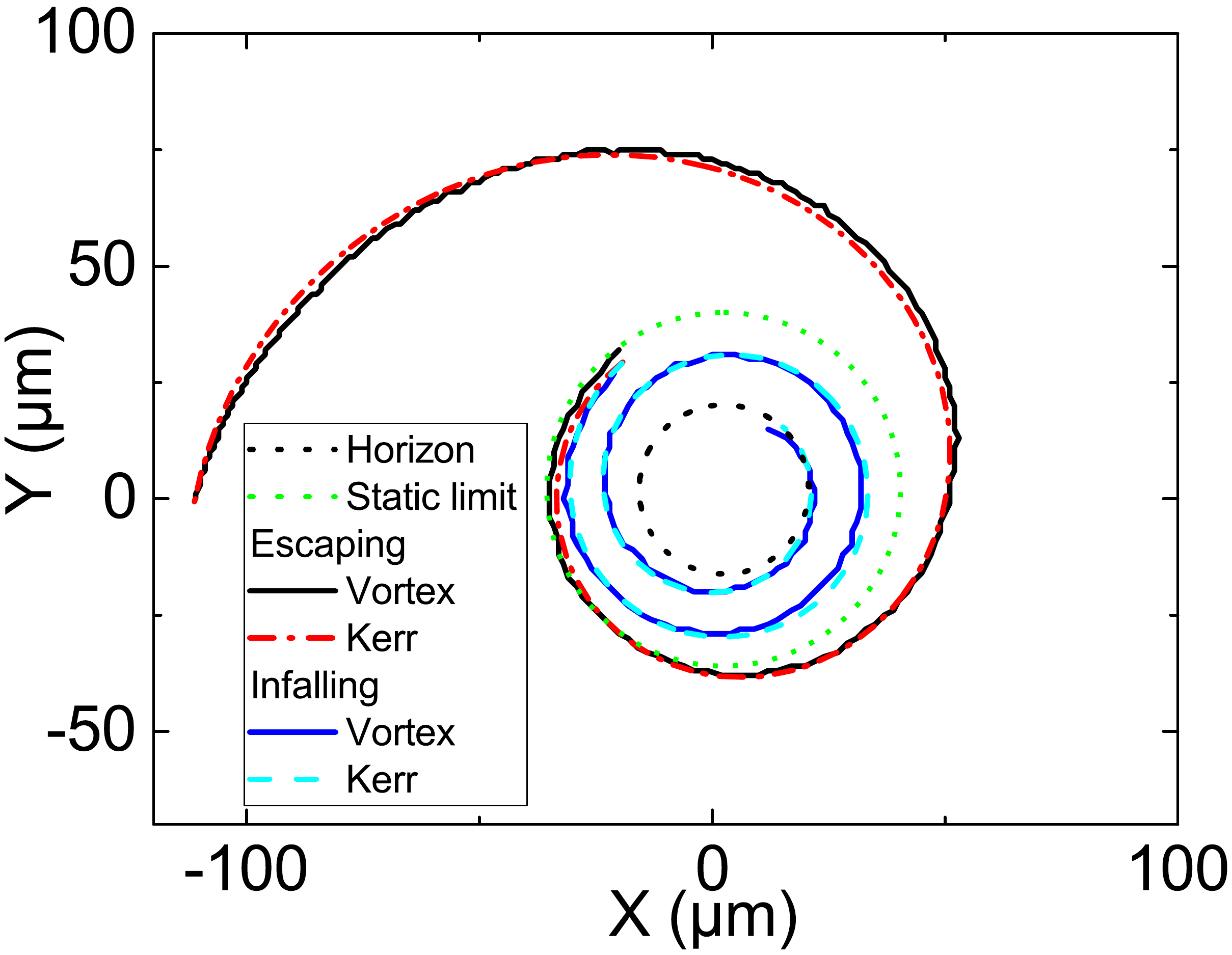}\\
  \caption{(Color online)  Trajectories of a quantum vortex (solid lines) and a massive particle (dashed/dash-dotted lines), in the vicinity of a Kerr black hole, exhibiting escaping (black, red) and infalling (blue, cyan) behaviors. Dotted lines: horizon (black), static limit (green).}
  \label{FigTraj}
\end{figure}

Interestingly, the trajectory of the remaining vortex can be well described by the time-like geodesics of the Kerr metric. We use the Hamiltonian formulation of the Kerr geodesic motion, providing a better numerical stability \cite{Levin2008}, where the equations of motion, relevant for the equatorial plane, read:
\begin{eqnarray}
\label{dyngeo}
\dot{r}&=&\frac{\Delta}{\Sigma}p_r\\
\dot{p}_r&=&-\left(\frac{\Delta}{2\Sigma}\right)'p_r^2+\left(\frac{R}{2\Delta\Sigma}\right)'\\
\dot{\phi}&=&-\frac{1}{2\Delta\Sigma}\frac{\partial}{\partial L}R
\end{eqnarray}
where prime denotes a partial derivative over $r$, with $\Sigma=r^2$, $\Delta=r^2-2Mr+a^2$,  $R=P^2-\Delta(r^2+(L-aE)^2)$, and $P=E(r^2+a^2)-aL$. The integrals of motion for the test particle are its energy $E$ and angular momentum $L$ (fitting parameter). The comparison of the numerical simulations of the vortex propagation with the Gross-Pitaevskii equation \eqref{GPE} (solid lines) with the time-like geodesic trajectories \eqref{dyngeo} (dashed/dash-dotted lines) is shown in Fig.~\ref{FigTraj} (the system size is larger than in Fig. \ref{snapshots}). A good agreement is obtained for both escaping and infalling trajectories, differing by the initial position $r$. 

The agreement of analog and real geodesic trajectories in the vicinity of the black hole is explained by the fact that the motion is mostly determined by the divergent metric term $g_{rr}$, behaving as $g_{rr}\sim (r-r_H)^{-1}$ for both the original Kerr metric Eq.~\eqref{Kerr} in general relativity and for the condensate Kerr metric Eq.~\eqref{Kcm}. For small rotation, one can estimate the effective mass of the analog black hole as $M\sim\xi\zeta$ \cite{suppl}. The vortex-vortex interaction beyond the metric (analog of electromagnetic interaction) is negligible for the test vortex for $\sqrt{\nu^2+\zeta^2}\gg 1$. We stress that at large distances, the condensate metric gives an effective potential $U_{cond}\sim -1/r^2$ \emph{different} from that of the Kerr metric $U_{Kerr}\sim -1/r$, leading to the deviation of the trajectory from the predictions of general relativity \cite{suppl}. Moreover, the vortex interaction with the wall becomes dominant at $R-r\approx\xi$. However, the condensate metric correctly reproduces the behavior in the immediate vicinity of the horizon, as shown above, confirming that an analog metric can be sufficiently close to the Kerr metric at least in the most important region for the simulation of Kerr black holes.

Although we dealt with a quantum fluid, similar phenomena could also be observable to some extent in classical fluids \cite{Euve2016,Torres2017}. We stress that our simulations in the mean-field approximation neglect the quantum fluctuations. However, these fluctuations represent one of the most interesting features of the analog systems, which make the scales of quantum mechanics and general relativity comparable. The natural outlook of the present work is therefore the study of the effect of quantum fluctuations (controlled via the condensate density \cite{suppl}) and the elucidation of quantum effects in analog gravity experiments by comparison with mean-field predictions for the development of quantum gravity \cite{Barcelo2005}. Another important direction could be the study of effective electrodynamics in strongly curved spacetimes, with vortices and bogolons playing the role of charges and photons. Finally, our work makes an important step towards self-consistent analog gravity: the metric in our case is not completely fixed externally, but depends on the matter and energy (vortex) distribution in the system.

To conclude, we have shown that a Kerr black hole with a quantized angular momentum can be created in a condensate in presence of a localized particle decay, that the quantum vortices as test particles close to such analog black hole follow  the time-like geodesics given by the general relativity for the Kerr metric. This configuration therefore represents a unique possibility to observe experimentally the propagation of massive particles along strongly non-Newtonian geodesics, far beyond the small relativistic corrections of the Mercury orbit. Finally, we have demonstrated the possibility of the analog Penrose effect, extracting the rotation energy of the Kerr black hole.

\begin{acknowledgments}
We acknowledge the support of the project "Quantum Fluids of Light"  (ANR-16-CE30-0021), of the ANR Labex Ganex (ANR-11-LABX-0014), and of the ANR program "Investissements d'Avenir" through the IDEX-ISITE initiative 16-IDEX-0001 (CAP 20-25). D.D.S. acknowledges the support of IUF (Institut Universitaire de France). S.V.K. acknowledges the support from the Ministry of Education and Science of Russian Federation (Project 16.9790.2017).
\end{acknowledgments}

\bibliography{biblio}

\renewcommand{\thefigure}{S\arabic{figure}}
\setcounter{figure}{0}

\section{Supplemental Material}

In this Supplemental Material we provide more details on the analytical solution and the numerical simulations discussed in the main text. We estimate the maximal angular momentum of the black hole and find its effective mass. We discuss the effect of the quantum and thermal fluctuations, neglected in the mean-field approach. We also describe the Supplemental Movies.

\subsection{Numerical simulations: parameters and details}

We have performed the simulations using mesh sizes between $N\times N=2^{8}\times 2^{8}$ and  $N\times N=2^{10}\times 2^{10}$,  with $h=0.5~\mu$m step size. The time step was $dt=2\times 10^{-15}$~s. We used third-order Adams-Bashforth method for the time integration of the Gross-Pitaevskii equation, both with and without the relaxation term (see below). The Laplacian term was calculated via a double Fourier transform, in order to obtain an efficient parallelization on the Graphics Processing Unit.

The polariton density and the interaction constant were chosen to have the interaction energy $\mu=\alpha n_{\infty}=1$~meV. The potential $U$ has both real and imaginary parts:
\begin{equation}
U=U_c(\mathbf{r})-i\Gamma(\mathbf{r}),
\end{equation}
where the real part describes an etched cylindrical mesa of a large diameter $R=100~\mu$m (or larger for some calculations) providing the confinement of the condensate:
\begin{equation}
U_c(\mathbf{r})=U_0\Theta(r-R)
\end{equation}
where $\Theta$ is the Heaviside's function.

The imaginary part ensures the convergent polariton flow towards the center. It describes a localized defect increasing the particle annihilation rate (e.g. defect in the cavity mirrors or a micrometric size metallic droplet on their surface):
\begin{equation}
\label{eq_gamma}
\Gamma(\mathbf{r}) = \frac{\hbar}{2\tau}e^{-\frac{r^2}{2\sigma^2}}
\end{equation}
where $\sigma=6~\mu$m is the size of the defect. The effective decay rate for the analytical approximation $\zeta$ discussed in the main text can be approximated as the average value $\langle\psi|\Gamma(r)|\psi\rangle$. It plays a role of a fitting parameter for Fig. 1(a,b) of the main text. We stress that contrary to the vorticity $\nu$, the decay parameter $\zeta$ is not quantized and does not have to be an integer.

\subsection{Stationary solution of the Gross-Pitaevskii equation}

To find the stationary solution of the Gross-Pitaevskii equation numerically, we introduce the phenomenological damping term \cite{Choi1998}, proposed by Pitaevskii in 1958 to describe the energy relaxation \cite{Pitaevskii58}. This is similar to using imaginary time integration technique. The damped Gross-Pitaevskii equation reads:
\begin{equation}
i\hbar \frac{{\partial \psi }}{{\partial t}} =\left(1-i\Lambda\right)\left( - \frac{{{\hbar ^2}}}{{2m}}\Delta \psi  + \alpha {\left| \psi  \right|^2}\psi +U\psi - \mu \psi \right)
\end{equation}
where $\Lambda$ is a dimensionless damping coefficient, which for reasonable calculation time and precision can range from $10^{-3}$ to $10^{-1}$.

A stationary solution $\psi_0$ of the non-damped Gross-Pitaevskii equation with an energy $\mu$ is also a solution of the damped equation: the right part of the equation simply vanishes and the presence of $\Lambda$ does not change anything. Moreover, any perturbations to the stationary solution $\psi_0$ increasing its energy tend to decay, and their decay rate is proportional to their energy deviation from $\mu$. This procedure conserves zeros of the wave function, and therefore allows to find stationary solutions different from the ground state, starting from an appropriate initial wavefunction. To improve convergence, we start with the initial wavefunction
\begin{equation}
\psi(r,\phi)=\sqrt{n_{\infty}}\tanh{\frac{r}{\xi}}\Theta(R-r)\exp{i\nu\phi}
\end{equation}
The initial state with a single high-winding vortex is split into $\nu$ single-winding vortices, which remain inside the horizon (see also the Supplementary section on the Maximal angular momentum below).

The  wavefunction $\psi_0$ found by the above numerical procedure is then used as a stationary solution on which the perturbations are created by time-dependent potential pulses.

\subsection{Dynamical simulations of the Penrose effect}

In order to implement a stationary convergent flow, we maintain a constant particle density far from the central region. This is equivalent to experimentally realistic situation of non-resonant pumping with a ring-like profile, which maintains the condensate at a constant level.

To perturb the condensate, we use an extra term in the real part of the potential profile $\delta U(\mathbf{r},t)$ with a different shape:
\begin{itemize}
\item For Fig. 2, we used a Gaussian-shaped potential pulse strongly elongated along X. The pulse is also Gaussian in time:
\begin{equation}
\delta U(\mathbf{r},t)=\delta U_0e^{-\frac{(x-x_0)^2}{2\sigma_x^2}}e^{-\frac{(y-y_0)^2}{2\sigma_y^2}}e^{-\frac{(t-t_0)^2}{2\sigma_t^2}}
\end{equation}
where $\sigma_x=30~\mu$m, $\sigma_y=3~\mu$m, $\sigma_t=0.5$~ps. The size of the perturbation along Y has to be large enough ($\sigma_y\gg\xi$) in order not to involve large wave vector bogolons, which propagate faster than the speed of sound $c$. At the same time, it has to remain small enough in order to remain along a single azimuthal direction. The amplitude here was $\delta U_0=0.1$~meV, much smaller than the interaction energy $\alpha n_{\infty}$
\item For Fig. 3, the size of the pulse has to be comparable with the size of a vortex pair $2\xi$:
\begin{equation}
\delta U(\mathbf{r},t)=\delta U_0e^{-\frac{(x-x_0)^2}{2\sigma_x^2}}e^{-\frac{(y-y_0)^2}{2\sigma_y^2}}e^{-\frac{(t-t_0)^2}{2\sigma_t^2}}
\end{equation}
with $\sigma_x=3~\mu$m, $\sigma_y=1~\mu$m, $\sigma_t=0.5$~ps. The orientation of this slightly elongated density minimum influences the splitting and the trajectories of the vortex and anti-vortex. We have kept this orientation  along the Y axis, in order to have a limited number of variable parameters in the system.
\end{itemize}

\subsection{Maximal angular momentum of the Black Hole}

As in general relativity, we find that the maximal angular momentum of the black hole analog is limited by its mass.

In general relativity, the maximal allowed value is $a_{max}/M=1$. One can rewrite this condition using the irreducible black hole mass (that of a non-rotating black hole), which we mark $M_0$: $a_{max}=\sqrt{2}M_0$.

To find the effective mass of the analog black hole in a condensate, 
we begin by comparing the dominant term of the metric for a non-rotating BH in general relativity and in a condensate. In general relativity:
\begin{equation}
g_{rr}=\frac{2M_0}{r-r_H}
\end{equation}
while in the condensate
\begin{equation}
g_{rr}=\left(1-\frac{v_r^2}{c^2}\right)^{-1} =\frac{1}{2}\frac{r_H}{r-r_H}
\end{equation}
which gives
\begin{equation}
M_0=\frac{1}{4}r_H=\frac{1+\sqrt{3}}{4\sqrt{2}}\xi\zeta\approx 0.5\xi\zeta 
\end{equation}
Therefore, the radius of the horizon is a good estimate for the analog black hole mass:
\begin{equation}
M\sim r_H
\end{equation}

The maximal number of vortices $\nu_{max}$ and thus the maximal angular momentum of a Kerr black hole analog can be estimated as follows. When $\nu$ is increased, the centrifugal force expels vortices towards the horizon. This behavior is different from that of trapped rotating condensates, where vortices are usually forming a lattice \cite{Ketterle2001}. Here, it is rather a chain of vortices which is formed \cite{Boulier2015} because of the effective energy profile. If $\nu_{max}$ vortices are located along the horizon, the localization length for each of them can be estimated as
\begin{equation}
\xi'= \frac{2\pi r_H}{\nu_{max}}
\end{equation}
The energy barrier which prevents vortices from escaping the black hole is given by the interaction energy $\mu=\alpha n$. This sets the condition for $\xi'$:
\begin{equation}
\frac{\hbar^2}{2m}\left(\frac{2\pi}{\xi'}\right)^2=\alpha n
\end{equation}
which allows to express $\xi'$ using the healing length $\xi$: $\xi'=2\pi\xi$. We can then find $\nu_{max}$:
\begin{equation}
\nu_{max}=\frac{r_H}{\xi}
\end{equation}
given in the main text. Expressing the angular momentum $a$ in the same natural units as the black hole mass $M$: $a_{max}=\nu_{max}\xi=r_H$, we obtain the  maximal angular momentum of the analog Kerr black hole:
\begin{equation}
\frac{a_{max}}{M}\sim 1
\end{equation}

This is confirmed by the numerics. For the same parameters of localized decay as in Figs. 2, 3 of the main text, the numerical result for the maximal possible number of vortices is $\nu=16$, which is indeed very close to the estimate $\nu_{max}=r_H/\xi=18$.

\subsection{Details of the Penrose process}
In the Penrose process, a particle $p$ entering the ergosphere separates into two particles $p'$ and $p''$, with $E_{p''}<0$, which allows the particle $p'$ to have an energy higher than the initial one: $E_{p'}>E_{p}$ in spite of the energy conservation $E_p=E_{p'}+E_{p''}$.

Here, we discuss the microscopic details of the separation of the initial density minimum into a vortex pair. We explain how the vortex-antivortex interaction allows the antivortex to enter a negative energy state and the vortex to gain a positive energy in order to escape the black hole.

\begin{figure}[tbp]
  \includegraphics[width=1\linewidth]{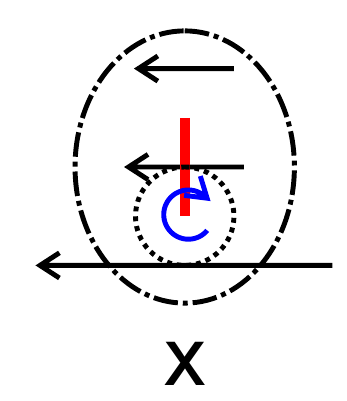}\\
  \caption{Scheme of the separation of a density minimum into a pair of vortices.}
  \label{sch}
\end{figure}

The density minimum created by a potential pulse could just disappear if the condensate was stationary. It is the gradient of the velocity of the flow close to the black hole which leads to the separation of this minimum into a vortex-antivortex pair. Indeed, the density minimum created in the condensate is deep enough to contain a zero-density line with undetermined phase. This line can spawn a pair of vortices if the conditions are favorable, which is indeed the case thanks to the velocity gradient, as shown in Fig.~\ref{sch}. We consider first the extreme point of the zero-density line (red line in the figure), closest to the center of the black hole (marked by a cross). The velocity circulation around this point (along the dashed circle) would have a tendency to become nonzero, if possible, and this can be realized by the formation of a vortex rotating in the direction opposite to that of the black hole (which we call an anti-vortex). Since the circulation around the dash-dotted line must remain zero (as it was initially), a vortex appears at the other end of the line, where the velocity gradient is much lower. In the end, it is this velocity gradient which breaks the symmetry and makes an anti-vortex appear closer to the black hole than the vortex.

\begin{figure}[tbp]
  \includegraphics[width=1\linewidth]{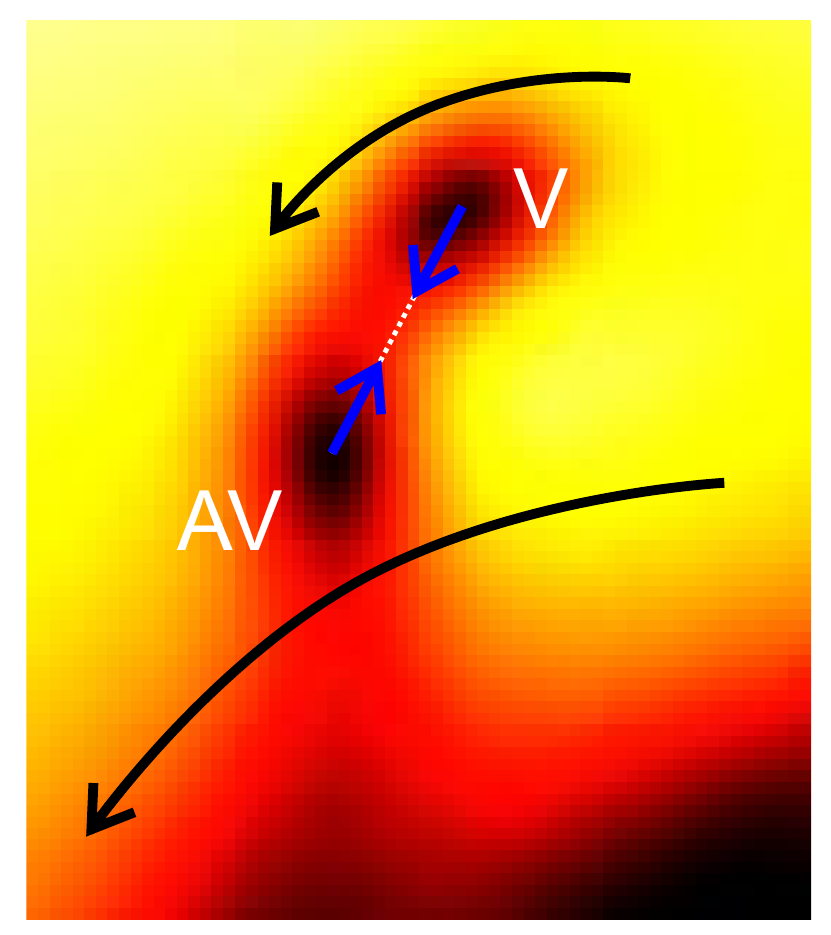}\\
  \caption{Vortex interaction bringing AV into a negative energy state and increasing the positive energy of V.}
  \label{sch2}
\end{figure}

Once the vortices are formed, it is the vortex-vortex "electromagnetic" interaction that explains the fact that the antivortex is brought into the negative energy state (Fig.~\ref{sch2}). Indeed, the vortices with opposite rotation are exhibiting a mutual attraction, as shown by the blue arrows along the white dashed line connecting their centers. In a stationary condensate, such attraction leads to the mutual annihilation of the vortex and the antivortex in the pair, but here it is not sufficient to overcome completely the attraction of the black hole. This interaction, as can be seen from the scheme, slows down the antivortex (AV) with respect to the flow while at the same time accelerating the vortex (V). This is the essential element of the analog Penrose process in our system. Indeed, close to the horizon, \emph{any} state rotating slower than the local frame has a negative energy \cite{Taylor} (which simply means that an energy higher than $mc^2$ is necessary to get the object to infinity). Therefore, the antivortex, being slowed down, gets into the state $E_{AV}<0$, and the vortex, being sped up, increases its energy (conserving the total).

We have checked the velocity of the vortices from our numerical simulations and found that at the antivortex position the azimuthal velocity $v_{\phi}=1.6~\mu$m/ps, whereas the antivortex velocity is $v_{AV}=1.4~\mu$m/ps$<v_{\phi}$. So, the velocity of the antivortex is smaller than that of the flow, meaning that it is indeed in the negative energy state.

We stress that this "electromagnetic" interaction is not, and is not supposed to be, described by the metric, exactly like the mechanism which separates the particle $p$ into $p'$ and $p''$ in the original Penrose process. It strongly depends on the V-AV distance and only plays a role at the first moments, whereas after the separation, the free propagation of the vortex becomes well described by the metric.

\subsection{Vortex trajectories and time-like geodesics of the Kerr metric}

In this section, we discuss the calculations linked with Fig.~4 of the main text, which compares the vortex trajectories with the time-like geodesics of the Kerr metric.

First of all, to avoid any misunderstanding, we would like to stress once again that the acoustic metric governing the vortices and the Kerr metric \emph{are} different, and one cannot reproduce the Kerr metric in the \emph{entire} infinite space using a condensate. Still, one of our important results is that they are essentially equivalent for a certain range of distances in the vicinity of the horizon, where the $g_{rr}$ components of both metrics exhibit identical scaling.

For the calculation of the geodesic trajectories of the Kerr metric, we start by estimating the parameters of the black hole. To increase the size of the ergosphere and facilitate the observation of the Penrose process, we usually put the maximal possible number of vortices into the black hole, which corresponds to $a/M=1$ in general relativity (see above). The value of $M$ is chosen to correspond to the position of the horizon and of the static limit, known from the analytical wavefunction and numerical simulations. The fitting parameters are the angular momentum of the particle $L$ and its initial radial momentum $p_r|_{t=0}$. The energy of the particle can be considered as fixed, because for any physically allowed values of the initial momentum determined by $L$ and $p_r|_{t=0}$, one can choose a particle rest mass $m_0$ to keep the energy constant. The values of the fitting parameters change in agreement to the observed behavior of the vortex, as a function of the initial position of the density minimum. When the vortex is created closer to the black hole, it feels a higher attraction and higher azimuthal dragging, and therefore acquires a higher radial momentum  $p_r|_{t=0}$ and higher angular momentum $L$ during the Penrose process. The values used to fit Fig.~4 of the main text are $p_r=-0.01$, $L=3.19$ for the escaping trajectory and $p_r=-0.4$, $L=4.0$ for the infalling trajectory.

The angular momentum of the black hole $a$ changes during the Penrose process, but we consider the vortex trajectory only for its free propagation part, after the end of the Penrose process, when the anti-vortex has already been annihilated. During this free propagation part the angular momentum of the black hole does not change, and considering it as constant is a good approximation. Moreover, since $\nu\gg 1$, $\nu-1\approx\nu$, and it is still reasonable to take $a/M=1$ even after the annihilation of one of the vortices of the black hole.

\begin{figure}[tbp]
  \includegraphics[width=1\linewidth]{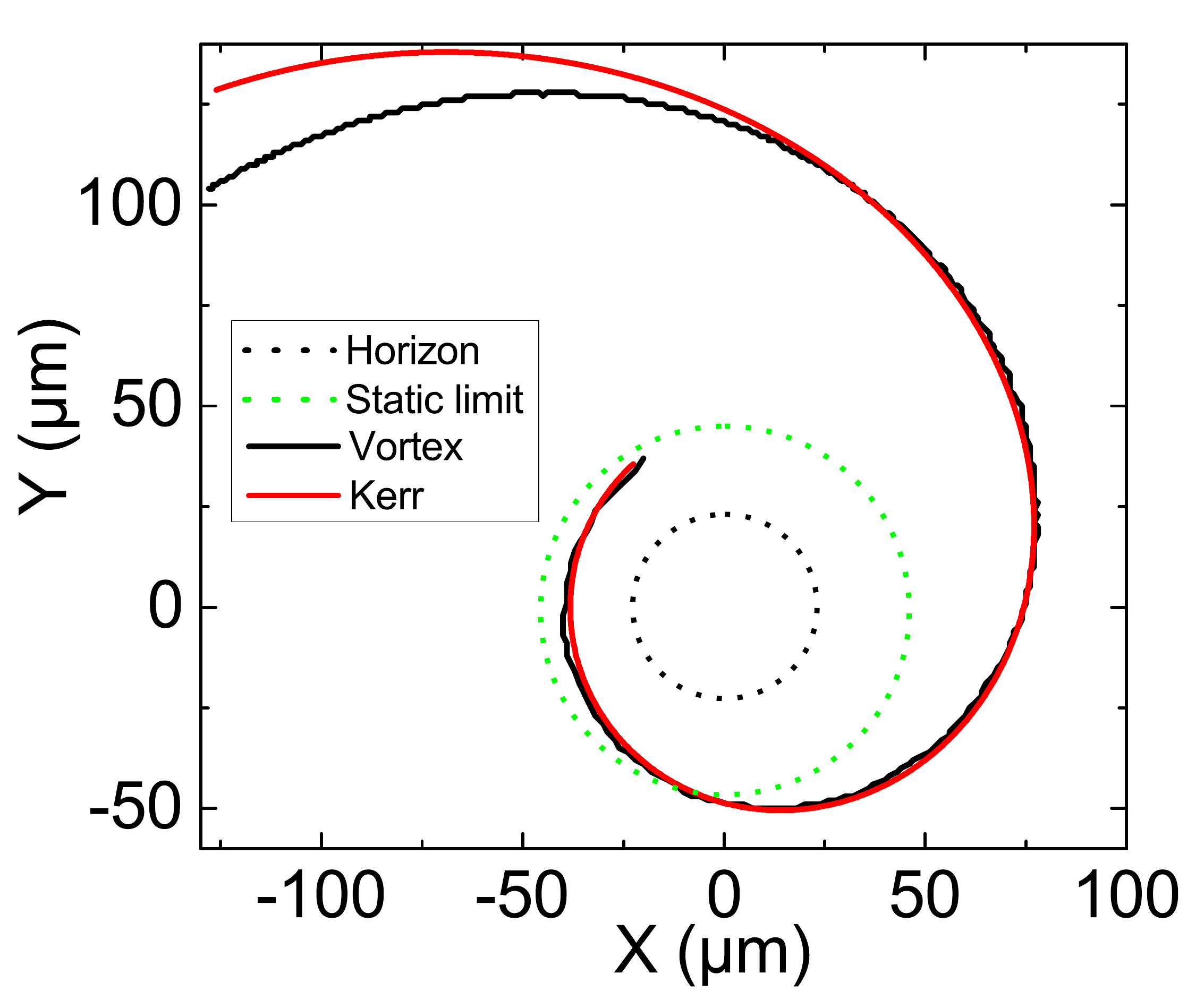}\\
  \caption{Comparison of the vortex trajectory and the Kerr time-like geodesic at large distances.}
  \label{f4sup}
\end{figure}

To illustrate the differences between the true and the analogue Kerr metrics, we made a larger-scale calculation, which shows that while the Kerr geodesic fits the vortex trajectory very well at short distances, the two diverge at larger $r$. Indeed, as stated in main text and as can be seen from the comparison of the two metric, the long-range potential decays as $-1/r$ in general relativity and as $-1/r^2$ in the condensate. A faster decay in the condensate means that the characteristic distances are reduced: for the same kinetic energy, the vortex goes less far away than a particle in the true Kerr metric. This is what is observed in Fig.~\ref{f4sup}: the Kerr geodesic (red line) fits the vortex trajectory (black line) very well at short distances, while at larger scales the vortex starts to lose the kinetic energy faster.

\subsection{Phase of the wavefunction}

\begin{figure}[tbp]
  \includegraphics[width=1\linewidth]{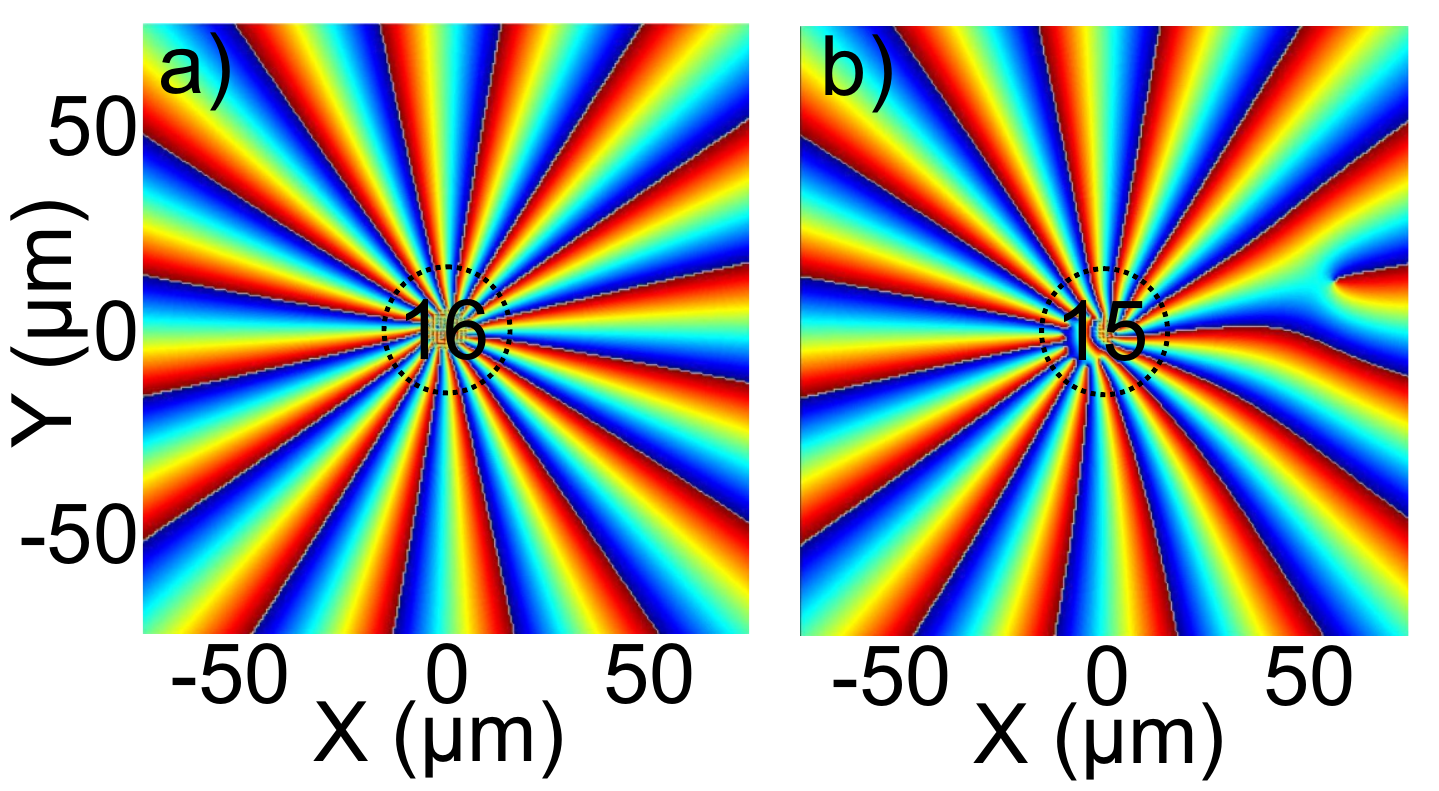}\\
  \caption{Phase patterns for a) stationary analog Kerr black hole solution with $\nu=16$ b) $\nu=15$ Kerr black hole after Penrose process. The 16th vortex is visible on the right.}
  \label{figS1}
\end{figure}

To confirm that the angular momentum of the black hole analog is indeed decreasing, we show the phase of the condensate wavefunction before and after the Penrose process takes place. In Fig.~\ref{figS1} we see that the number of vortices in the black hole is reduced from $\nu=16$ (panel a) to $\nu=15$ as a result of the Penrose process, as described in the main text. Moreover, we can see that the position of the vortices in the black hole is perturbed, and this perturbation remains visible for a long time, which might be a possible signature of the excitation of the quasibound states \cite{Patrick2018}.

\subsection{Mean-field theory and fluctuations}

In the present work, we neglect the quantum and thermal fluctuations of the condensate, limiting the consideration to the mean-field approach. In 2D systems at 0 K quantum fluctuations lead to a depletion of the condensate determined by the interactions \cite{Pitaevskii}. In 2D, the quantum depletion can be estimated from the one-body density matrix, as the relative difference between its value at zero distance (local correlations determined by both the condensed and the excited fractions) and at infinity (determined only by the condensate, by the definition of the long-range order):
\begin{equation}
\frac{n^{(1)}(0)-n^{(1)}(\infty)}{n^{(1)}(0)}\sim\frac{1}{n\xi^2} \sim\frac{\alpha m}{\hbar^2}
\end{equation}
We see that the relative importance of quantum depletion can be controlled via the interactions or the particle mass, which can be tuned by using the Feshbach resonance (in atomic condensates) and the detuning (in polariton condensates). On the one hand, this allows to reproduce the mean-field results obtained in the present manuscript by reducing the interactions and the mass. On the other hand, increasing the effect of the quantum fluctuations opens the way towards analog quantum gravity. Comparing the results obtained in the two regimes allows to make clear the effects of a fluctuating metric.

At non-zero temperatures, thermal fluctuations induce a power-law decay of the coherence:
\begin{equation}
n^{(1)}(s)\sim\left(\frac{s_T}{s}\right)^\nu
\end{equation}
with
\begin{equation}
\nu=\frac{k_B Tm}{2\pi\hbar^2n_s}
\end{equation}
where $n_s$ is the superfluid density. For polaritons at 10 K (typical for GaAs cavities providing the best quality), $s_T\approx1.3~\mu$m and $\nu\approx 5\times 10^{-4}$, which gives a very slow decay of the condensate coherence with respect to all other possible sources. We therefore expect that thermal fluctuations should not cause too much problems for the observation of the effects discussed in the main text.

\section{Supplemental movies}
In this section, we discuss the supplemental movie files.

\begin{itemize}
\item \textsf{penrose.avi} - The movie shows all the stages of the Penrose effect, similar to the snapshots of Fig.~3 of the main text. A density dip is created by a localized pulsed potential, this density dip is dragged into the ergosphere, where it splits into a vortex-antivortex pair. The antivortex falls into the black hole and annihilates a vortex inside it, thus reducing its angular momentum, while the vortex of the pair escapes to infinity.
\item \textsf{escaping.avi} - The movie shows the trajectory of the vortex from the vortex-antivortex pair after its formation. The vortex is sufficiently far from the black hole to be able to escape to infinity. The position of the vortex (detected from velocity curl) is marked by a cross. The trajectory extracted from this movie is plotted in Fig. 4 of the main text as a black solid line.
\item \textsf{infalling.avi} - The movie shows the trajectory of the vortex from the vortex-antivortex pair after its formation. The vortex is a little bit closer to the black hole than in \textsf{escaping.avi} and falls inside. The position of the vortex (detected from velocity curl) is marked by a cross. The trajectory extracted from this movie is plotted in Fig. 4 of the main text as a blue solid line.
\item \textsf{single\_infalling.avi} - The movie shows a vortex created at $t=0$ as a part of the initial solution $\psi$ at the same distance from the black hole, as the escaping vortex formed from a pair. Here (and, actually, with any starting distance), the vortex falls into the black hole, confirming that without the extra energy provided by the generation of the antivortex with negative energy, the vortex cannot escape to infinity.
\end{itemize}

\end{document}